\documentclass[11pt]{article}
\usepackage{amsmath,amsthm,amssymb,amsfonts}
\usepackage[a4paper]{geometry}
\usepackage{a4wide}
\usepackage{graphicx}
\usepackage{booktabs}
\usepackage{rotating}
\usepackage{colortbl}
\usepackage{color}

\usepackage{textcomp}
\usepackage{url}
\usepackage{color}
\usepackage{xcolor}
\begin{document}

\makeatletter
%%%%%------- preprint style ----%%%%%%%%%%%%%%%%%%%%
\@addtoreset{equation}{section}
\def\theequation{\thesection.\arabic{equation}}
\def\@maketitle{\newpage
 \null
 {\normalsize \tt \begin{flushright} 
  \begin{tabular}[t]{l} \@date  
  \end{tabular}
 \end{flushright}}
 \begin{center} 
 \vskip 2em
 {\LARGE \@title \par} \vskip 1.5em {\large \lineskip .5em \begin{tabular}[t]{c}\@author 
 \end{tabular}\par} 
 \end{center}
 \par
 \vskip 1.5em} 
\makeatother
%%%%%%%%%%%%%%%%%%%%%%%%%
\topmargin=-1cm
\oddsidemargin=1.5cm
\evensidemargin=-.0cm
\textwidth=15.5cm
\textheight=22cm
%\renewcommand{\baselinestretch}{1.5}
%%%%%%%%%%%%%%%%%%%%%%%%
\setlength{\baselineskip}{16pt}
\title{Kerr/Fluid Duality and Singularity of Solutions to the Fluid Equation
%Caustics of Null Geodesics on a  Horizon
}

\author{
Ippei~{\sc Fujisawa} and
Ryuichi~{\sc Nakayama}
       \\[1cm]
{\small
    Division of Physics, Graduate School of Science,} \\
{\small
           Hokkaido University, Sapporo 060-0810, Japan}
}
\date{
EPHOU-15-017  \\
%November 2015
}
% 
%\begin{titlepage}
% 
\maketitle

\begin{abstract} 
An equation for a viscous incompressible fluid on a spheroidal surface which is dual to the perturbation around the near-near  horizon extreme Kerr (near-NHEK) black hole is derived.  It is also shown that an expansion scalar $\theta$ of a congruence of null geodesics on the perturbed horizon of the perturbed near-NHEK spacetime, which is dual to a viscous incompressible fluid, is not in general positive semi-definite, even if  initial conditions on the velocity are smooth. Unless initial conditions are elaborated,  caustics of null congruence will occur on the perturbed horizon in the future. A similar result is obtained for a perturbed Schwarzschild black hole spacetime which is dual to a viscous incompressible fluid on $S^2$. An initial condition that $\theta$ be positive semi-definite at any point on $S^2$ is a necessary condition for the existence  of smooth solutions to incompressible Navier-Stokes (NS) equation on $S^2$. 
\end{abstract}
%\end{titlepage}
\newpage
\setlength{\baselineskip}{18pt}

%%%%%%%%%%%%%%%%%%%%%%%%%%%%%%%%%%%%%%%%%%%%%%%%%%%%%%%%%%%%%%%%%%%%%
\newcommand{\bm}[1]{\mbox{\boldmath $#1$}}
%%%%%%%%%%%%%%%%%%%%%%%%%%%%%%%%%%%%%%%%%%%%%%%%%%%%%%%%%%%%%%%%%%%%

%\begin{document}
%\maketitle

\section{Introduction}
\hspace{5mm}
Fluid/Gravity correspondence has a long history starting from \cite{Damour} and \cite{PriceThorne}. In the context of AdS/CFT this subject has been extensively studied.\cite{Policastro}\cite{Wadia} In the case of asymptotically flat spacetimes a prescription was developed in \cite{Str4} and the incompressible Navier-Stokes (NS) equation on a plane which is dual to Rindler space\cite{Str5} and that  on a sphere which is dual to Schwarzschild black hole\cite{Str1}\cite{Taylor} are derived.  Furthermore Petrov type I condition on the hypersurface geometry reduces the Einstein equation to the NS equation.\cite{Str6} To our knowledge, however,  an equation of fluid mechanics which is dual to a more realistic black hole, a rotating one in an asymptotically flat spacetime has not been obtained until now.  

In this paper we will derive an extension of an incompressible Navier-Stokes (NS) equation to that which is dual to the Kerr black hole solution in four-dimensional asymptotically flat spacetime.
For simplicity, we will consider the near-near horizon extreme Kerr (near-NHEK) limit\cite{Str2} which is introduced in the context of Kerr/CFT correspondence \cite{Str3}\cite{Cargese}.
This problem is interesting from the point of view of symmetry. Kerr black hole has only axial symmetry and 
the horizon of the Kerr Black hole is not a round sphere, but a distorted, spheroidal surface. This reduces the symmetry of the fluid equation, and it makes it difficult to infer the form of the fluid equation solely from symmetry considerations.  

Another subject of this paper is the expansion scalar $\theta$ of the null congruence of the perturbed event horizon of the 
near-NHEK spacetime. In this paper this perturbed event horizon ${\cal H}$ is defined by the condition that ${\cal H}$ coincides with the stationary event horizon in the absence of gravitational perturbation, and ${\cal H}$ is null. This definition was also adopted for example in \cite{BHL}. This ${\cal H}$ is generated by null geodesics. In the case of the spacetime we study in this paper such a null surface ${\cal H}$ is uniquely determined by the $\lambda$ expansion which is introduced in the derivation of the incompressible NS equation.
We will compute $\theta$ to the leading non-trivial order of the $\lambda$ expansion and find that it is not necessarily positive semi-definite. 
If $\theta$ is negative at some point the congruence of null geodesics will end with caustics in finite time due to the focusing equation. At the caustics we have $\theta=-\infty$ and the velocity $v_i$ of the fluid becomes singular, since $\theta$ is a polynomial of $v_i$ and its derivatives.   This means that the solution to the incompressible NS equation has singularities, even if it is smooth in the past. To avoid such a singularity we need to impose an initial condition that $\theta$ be non-negative anywhere on ${\cal H}$ at some time $\tau=\tau_0$. The calculation of the expansion scalar is carried out for both near-NHEK  and Schwarzschild black hole. This result will be relevant to the problem\cite{CMI} of the existence and smoothness of solutions to the incompressible NS equation in spaces with topology different from a plane.\footnote{  It can be shown by similar calculation that $\theta$ is positive semi-definite in the case of a two- and three-dimensional planar horizons in Rindler space. }

By the assumption of cosmic censorship\cite{Penrose} it is considered that with a smooth initial condition there are no singularites on and outside the event horizon in the future. 
It is known that the entropy of a black hole is related to the area of the event horizon. The area of the event horizon and the entropy will not decrease.\cite{2ndlaw} In the non-stationary case, there are some proposals for definition of the horizons.\cite{Ashtekar2} In the previous paragraph we defined the event horizon of a perturbed spacetime as a null surface ${\cal H}$ which agrees with the event horizon of the unperturbed one when the perturbation is switched off. Then, if $\theta$ were negative at some point on ${\cal H}$, ${\cal H}$ would not be  a true event horizon. The true event horizon will be outside ${\cal H}$. 
However, even if this is the case, the above argument for the singularity of the solution to the fluid equation still applies, since the argument  is valid on some null surface ${\cal H}$. 
To the contrary, when $\theta$ is non-negative all over ${\cal H}$, ${\cal H}$ might as well be regarded as an event horizon.

This paper is organized as follows. In sec.2 procedure for constructing fluid mechanics dual to asymptotically flat spacetime\cite{Str5}\cite{Str1} is reviewed. Then the limit of near horizon extreme Kerr (NHEK) spacetime and the near-NHEK spacetime is explained. In sec. 3 an incompressible NS equation on spheroidal surface which is dual to near-NHEK is derived. This equation contains several extra terms. In sec. 4 the expansion scalar $\theta$ for the congruence of the null geodesics on the perturbed event horizon ${\cal H}$ is computed.  Discussions are given in sec. 5. In appendix A quantities related to the un-perturved horizon, a spheroidal $S^2$ is collected. In appendix B the perturbation part of the metric which is dual to the fluid is presented. In appendix C we report on some results for fluid equation which is dual to a perturbed Schwarzschild black hole and which was obtained by using the same prescription as for near-NHEK spacetime.

\section{Fluid/Gravity correspondence in asymptotically flat spacetime}
\hspace{5mm}
In this section we will recall some ingredients for deriving NS equation from Einstein gravity. We will restrict 
discussions to asymptotically-flat spacetimes. We start with some static or stationary black hole solution. 
An idea is to intorduce a cutoff surface $\Sigma_c$ at some fixed radius $r=r_c$ outside a black hole. We then impose some boundary condition for gravitational perturbations on $\Sigma_c$ and look for solutions which are non-singular at the horizon. Einstein equation has a solution space which is much larger than that of NS equation. By choosing suitable boundary conditions on $\Sigma_c$ it is possible to make some degrees of freedom remain which describe fluids on the horizon.  

In the case of flat Minkowski space, we choose in-going `Rindler coordinates',\cite{Str5} 
\begin{equation}
ds_\mathrm{Rindler}^2= -\hat{r} d\hat{t}^2+2d\hat{t} d\hat{r}+(dx^i)^2, \label{Rindler}
\end{equation}
and introduce a dimensionless parameter $\lambda$ and change coordinates $\hat{r},\hat{t} \rightarrow , t$ with  
\begin{equation}
\hat{r}= r_c+\lambda \, r, \qquad \hat{t}=\frac{t}{\lambda}.
\end{equation}
Upon introducing metric perturbations, we demand that the induced metric on the cut-off surface $\Sigma_c$ be flat. 
The cutoff surface approaches  the horizon by taking a limit $\lambda \rightarrow 0$. Within a power series expansion  around $\lambda=0$ we determine the metric components of the perturbations in terms of the velocity field $v^i(t,x)$ and pressure $P(t,x)$ in such a way that the full metric including the perturbations satisfies Einstein equation as far as $v^i$ and $P$ satisfy NS equation.  In taking the limit $\lambda \rightarrow 0$ the coordinate frame is infinitely accelerated and this allows us to probe the near-horizon dynamics. 
In this way the NS equation and the incompressibility condition are obtained. 
\begin{align}
& \partial_t v_i + v^j \partial_j v_i + \partial_i P - \nu \, \partial^2 v_i = 0, \label{NS_flat} \\
& \partial_i v^i = 0. \label{incomp_flat}
\end{align}
Here $\nu$ is the viscosity parameter. 
The above equations have scaling symmetry.
\begin{eqnarray}
v_i(t,x^i) &\rightarrow & \epsilon \, v_i(\epsilon^2 \, t, \epsilon \, x^i), \\
P(t,x^i) &\rightarrow& \epsilon^2 \, P(\epsilon^2 \, t, \epsilon \, x^i). \nonumber 
\end{eqnarray}
Here $\epsilon$ is a small constant parameter. 

In the case of Schwarzschild black hole, the incompressible NS equation 
\begin{align}
& \partial_t v_i + v^j \nabla_j v_i + \nabla_i P - \nu \, \left(  \nabla^2 v_i + R_{ij} v^j \right) = 0, \label{NS_Sch} \\
& \nabla_i v^i = 0. \label{incomp_Sch}
\end{align}
was obtained by a similar procedure.\cite{Str1}  Here, $R_{ij}$ is Ricci tensor on a round sphere and $\nabla_i$ is a covariant derivative associated with the metric on the sphere.
One notable difference is that it is not possible to impose a boundary condition such that the metric on $\Sigma_c$ be that of a direct product of time and a round sphere of radius $r_c$, but what can be done at most is that the induced metric on $\Sigma_c$ is conformal to such a fixed metric. 

In the above study symmetry principles are important to understand the final fluid equations. The equation (\ref{NS_flat}) is invariant under spacetime translation, rotation, scale transformation and parity.
The symmetries which originate from the isometries of the background spacetime restrict the form of the equation. In the background of Schwarzschild black hole, there are also enough symmetries to determine the form of the fluid equation.
 Actually, Killing vectors of round $S^2$ solve (\ref{incomp_Sch}). 

In the case of Kerr black hole, however, the spacetime does not have sufficient symmetries to determine the form of the fluid equation uniquely. 
Actually, rotation symmetry is reduced to only  $\partial_{\varphi}$. The horizon is not a round sphere, but a deformed, spheroidal surface. 
This suggests that the equation for the fluid on the spherical horizon also has only a rotation symmetry $\partial_{\varphi}$.
To the best of our knowledge, a form of NS equation or its generalization, which is dual to  such a curved spacetime without sufficient isometry has not been studied.
Hence we cannot predict {\it a priori} the exact form of the NS equation soley from the symmetry argument due to the lack of symmetries.
In this paper we will show that this fluid equation can be determined uniquely by requiring that the equation can be written in terms of covariant derivatives associated with the metric on the horizon. 

Another problem associated with the Kerr spacetime is the fact that it is more complicated than that of Schwarzschild.
It is known that the Kerr spacetime can be simplified by considering near horizon limit of the extremal Kerr or that of near extreme Kerr, and the correspondence with conformal field theories has been studied.  In this paper we will treat this limiting  case of 
Kerr spacetime. 

\subsection{NHEK and near-NHEK}
\hspace{5mm}
The metric of the Kerr black hole in Boyer-Lindquist coordinates $(\hat{r},\hat{t},\hat{\theta},\hat{\varphi})$ is given by
\begin{align}
ds_\mathrm{BL}^2 = & -\frac{\Delta}{\Sigma} \left( d\hat{t} - a \sin^2{\hat{\theta}}\, d\hat{\varphi} \right)^2
+ \frac{\sin^2{\hat{\theta}}}{\Sigma} \left( (\hat{r}^2+a^2) \, d\hat{\varphi}-a d\hat{t} \right)^2 
+ \frac{\Sigma}{\Delta} d\hat{r}^2 + \Sigma \, d\hat{\theta}^2  \label{BL}
\end{align}
where
\begin{align}
\Sigma(\hat{r}, \hat{\theta}) = \hat{r}^2 + a^2 \cos^2{\hat{\theta}}, \quad
\Delta(\hat{r}) = \hat{r}^2 - 2M\hat{r} + a^2. \label{SigmaDelta}
\end{align}
This metric has two parameters: the mass of the black hole $M$ and the angular momentum $J = a M$.
There is an event horizon at  $\hat{r} = r_+$, where $r_{\pm} \equiv M \pm \sqrt{M^2 - a^2}$ are solutions to  $\Delta (\hat{r})=0$. 
Due to the square root in the above definition for $r_{\pm}$, to avoid naked singularity, $a$ must satisfy an inequality $|a| \leq M$. 
The Hawking temperature, angular velocity of the horizon and entropy are 
\begin{eqnarray}
T_H &=& \frac{r_+-r_-}{8\pi M r_+}, \\
\Omega_H &=& \frac{a}{2Mr_+}, \\
S_{BH} &=& 2\pi M r_+.
\end{eqnarray}
We wish to remove the apparent singularity of (\ref{BL}) at the horizon, and we  change coordinates to in-going Kerr coordinates.
They are analogous to the Eddington-Finkelstein coordinates for Schwarzschild black hole  
which are appropriate for discussing perturbations around the horizon.
The necessary coordinate transformations and the resulting metric are, respectively,
\begin{equation}
dr = d\hat{r}, \quad
dt = d\hat{t} + (\hat{r}^2+ a^2) \frac{d\hat{r}}{\Delta(\hat{r})}, \quad
d\theta = d\hat{\theta}, \quad
d\varphi = d\hat{\varphi} + a \frac{d\hat{r}}{\Delta(\hat{r})}
\end{equation}
and
\begin{align}
ds_\mathrm{EF}^2 = & -\left( 1-\frac{2Mr}{\Sigma} \right) dt^2 + \Sigma \, d\theta^2 
+ \left(r^2+a^2 + \frac{2Mr}{\Sigma} a^2 \sin^2{\theta} \right) \sin^2{\theta} \, d\varphi^2 \nonumber \\
& +2dt dr -2a \sin^2{\theta} \, dr d\varphi - \frac{4Mr}{\Sigma} a \sin^2{\theta} \, dt d\varphi . \label{metEF}
\end{align}

The extreme Kerr has $r_+=r_-$, $T_H=0$. In this case the upper  bound for $J$ is saturated: $J=M^2$. 
We can blow up the near horizon region by the following replacement 
\begin{equation}
r \rightarrow M+\lambda M r, \quad t \to \frac{2M}{\lambda} t , \quad
\theta \to \theta, \quad
\phi \to \phi + \frac{t}{\lambda},\label{Extremechange}
\end{equation}
followed by a limit $\lambda \rightarrow 0$. The resulting metric is given by 
\begin{equation}
ds^2_\mathrm{NHEK}= 2J \Gamma(\theta) \, \Big(-r^2 dt^2+2dtdr+d\theta^2+\Lambda(\theta)^2(d\phi+rdt)^2\Big),
\end{equation}
where 
\begin{equation}
\Gamma(\theta) = \frac{1+\cos^2{\theta}}{2},\qquad \Lambda(\theta)=\frac{2\sin \theta}{1+\cos^2 \theta},
\end{equation}
and the associated geometry is the near horizon extremal Kerr (NHEK). \cite{
BardeenHorowitz}\cite{Str3}

While NHEK has vanishing Hawking temperature, its generalization with nonvanishing temperature, a near-NHEK spacetime was constructed.\cite{Str2} This is achieved by  taking the limit $\lambda \rightarrow 0$,  while keeping the associated temperature
\begin{equation}
T_R \equiv \frac{2 M T_H}{\lambda}
\label{righttemp}
\end{equation}
finite and non-zero.  This means that parameter $a$ is adjusted according to 
\begin{equation}
a=\frac{M\sqrt{1-4\pi T_R \lambda}}{1-2\pi T_R \lambda}.
\end{equation}

Now we make the following coordinate transformations to the metric (\ref{metEF}) 
\begin{equation}
r \to r_+ + \lambda \, r_+ \, r, \quad
t \to \frac{2M}{\lambda} t , \quad
\theta \to \theta, \quad
\phi \to \phi + \frac{t}{\lambda} \label{change}
\end{equation}
instead of the transformation (\ref{Extremechange}).

In the limit $\lambda \to 0$, metric (\ref{metEF}) becomes  
\begin{align}
ds_\mathrm{nNHEK}^2 = & \left( -\frac{16 M^2 \pi^2 T_R^2 (\Gamma-1)}{\Gamma}
+\frac{M^2(17-19 \Gamma + 6 \Gamma^2)}{4\Gamma} \left( 4\pi T_R \, r + r^2  \right) \right) dt^2 \notag \\
& + 4M^2 \Gamma dr dt 
- \frac{8M^2(\Gamma-1)}{\Gamma} \left(2 \pi T_R + r \right) dt d\varphi
+ 2M^2 \Gamma d\theta^2 - \frac{4M^2 (\Gamma-1)}{\Gamma} \, d\varphi^2.
\end{align}
This is called near-NHEK geometry.\cite{Str2} 

\section{Navier-Stokes equation dual to near-NHEK}
\subsection{Scale transfomation}
We will perform an extra rescaling of  the coordinates $r$ and $t$ (\ref{change}) to (\ref{metEF}) by
\begin{equation}
r = \lambda \, \rho, \quad
t = \frac{\tau}{\lambda} \label{scale}
\end{equation}
with the same parameter $\lambda$ as in the previous subsection. 
As $\lambda \rightarrow 0$, the coordinate frame is infinitely accelerated, and the near-horizon dynamics can be probed. 
The metric after the transformation (\ref{scale}) and expansion in power series of $\lambda$ is given by 
\begin{align}
ds^2 = & \frac{1}{\lambda^2} \frac{16M^2 \pi^2 T_R^2 (1-\Gamma)}{\Gamma} d\tau^2 \notag \\
&+\frac{1}{\lambda} \left( \left(-\frac{32M^2 \pi^3 T_R^3(1-4\Gamma + 3\Gamma^2)}{\Gamma^2}
-\frac{8M^2 \pi T_R (-2+2\Gamma+\Gamma^2)}{\Gamma} \rho \right) d\tau^2 \right. \notag \\
 &\left. ~~~~~~~~ + \frac{16M^2 \pi T_R (1-\Gamma)}{\Gamma} d\tau d\varphi \right) \notag \\
 & + 4M^2 \Gamma d\rho d\tau + 2M^2 \Gamma d\theta^2 + \frac{4M^2(1-\Gamma)}{\Gamma} d\varphi^2 \notag \\
& + \left( -\frac{16M^2 \pi^4 T_R^4 (-4+24\Gamma-49\Gamma^2 + 29\Gamma^3)}{\Gamma^3} \right. \notag \\
& ~~~~ -\frac{16M^2 \pi^2 T_R^2(3-10\Gamma+6\Gamma^2 + 2\Gamma^3)}{\Gamma^2}\rho
\left. - \frac{2M^2(-2+2\Gamma + \Gamma^2)}{\Gamma} \rho^2 \right) d\tau^2 \notag \\
& -\left( \frac{16M^2\pi^2 T_R^2(2-7\Gamma+5\Gamma^2)}{\Gamma^2} 
+ \frac{8M^2(-1+\Gamma)}{\Gamma} \rho \right) d\tau d\varphi  + \mathcal{O}(\lambda^2)           \label{met_-2}
\end{align}
and it turns out that a term with a power $\lambda^{-2}$ appears in front of $d\tau^2$. 
This $\lambda^{-2}$ order term can be removed by means of a shift in $\varphi$. 
\begin{equation}
\varphi \to \varphi - \frac{2 \pi T_R}{\lambda} \, \tau \label{phishift}
\end{equation}
After this shift  we obtain
\begin{align}
d\overline{s}^2 = & -\frac{8M^2\pi T_R \, \Gamma \, \rho}{\lambda} d\tau^2 
+  4M^2 \Gamma d\tau d\rho
\nonumber \\
& + \left( \frac{8M^2 \pi^4 T_R^4 \sin^2{\theta}}{\Gamma} 
- \frac{16M^2 \pi^2 T_R^2 \cos^2{\theta}}{\Gamma} \rho
- \frac{M^2 (3+28\cos{2\theta}+\cos{4\theta})}{16\Gamma} \rho^2
 \right) d\tau^2 \nonumber \\
& + \left(
 \frac{8M^2\pi^2 T_R^2 \sin^2{\theta}}{\Gamma} + \frac{4M^2 \sin^2{\theta}}{\Gamma} \rho
 \right) d\tau d\varphi 
 + 2M^2 \Gamma d\theta^2 +\frac{ 2M^2 \sin^2{\theta}}{\Gamma} d\varphi^2
 \nonumber \\
& + \lambda \Biggl( \Biggr.
 \left( \frac{16M^2 \pi^5 T_R^5 (2+\cos{2\theta})\sin^2{\theta}}{\Gamma^2}
 -\frac{32M^2\pi^3 T_R^3 \cos^4{\theta}}{\Gamma^2} \rho \right. 
 \nonumber \\
& \left. +\frac{M^2 \pi T_R (23+12 \cos{2\theta}-3\cos{4\theta})}{8\Gamma^2} \rho^2 \right) d\tau^2
+ 8M^2 \pi T_R d\tau d\rho 
\nonumber \\
&+ \left( \frac{16M^2 \pi^3 T_R^3 ( 5+3 \cos{2\theta} ) \sin^2{\theta}}{\Gamma^2}
+\frac{M^2 \pi T_R (35+28\cos{2\theta} + \cos{4\theta}) \sin^2{\theta}}{4\Gamma^2} \rho \right) d\tau d\varphi 
\nonumber \\
&+ 4M^2\pi T_R d\theta^2 + \frac{M^2\pi T_R \sin^2{2\theta}}{\Gamma^2}d\varphi^2 \Biggl. \Biggr) 
+ \mathcal{O}(\lambda^2). \label{met_bg}
\end{align}
We also present some components of the inverse metric up to $\mathcal{O}(\lambda^1)$.
\begin{gather}
g^{\rho\rho} = \frac{1}{\lambda} \frac{2\pi T_R \, \rho}{M^2 \Gamma} + \frac{\rho(-8\pi^2 T_R^2 + \Gamma \rho)}{2 M^2 \Gamma^2} + \mathcal{O}(\lambda^1), \quad g^{\rho\tau} = \frac{1}{2M^2 \Gamma} + \mathcal{O}(\lambda^1) \notag \\
g^{\rho\varphi} = - \frac{2\pi^2 T_R^2 + \rho}{2M^2 \Gamma} + \mathcal{O}(\lambda^1)
, \quad g^{\theta\theta} = \frac{1}{2M^2 \Gamma} + \mathcal{O}(\lambda^1)
, \quad g^{\varphi\varphi} = \frac{-1+2 \csc^2{\theta}}{4M^2} + \mathcal{O}(\lambda^1). \label{inv_bg}
\end{gather}
The leading order of $g^{\rho\rho}$ is also $\lambda^{-1}$ like that of  $g_{\tau\tau}$. 

We define $\Sigma_c$ as a hypersurface  located at $\rho=1$.
By taking a limit $\lambda \to 0$, $\Sigma_c$ approaches the event horizon $r=r_+$, which is 
a null surface. 
This is a distorted 2-sphere $S^2$, whose metric $\gamma_{ij}$ is given in (\ref{distortedsphere}). 
We attempted to impose a boundary condition on $\Sigma_c$ such that the induced metric on $\Sigma_c$ is 
(possibly conformal to) that of a semi-direct sum of  time and the distorted sphere.  It turns out, however, that this is not possible.

\subsection{Derivation of the Navier-Stokes equation}
\hspace{5mm}
We consider perturbation around the Kerr background, and we wish to find a fluid equation which is an extention of NS equation with possibly some appropriate extra terms.
The metric with perturbations will be denoted as
\begin{align}
ds^2 = d\overline{s}^2 + g_{\mu\nu}^\mathrm{(pert.)} dx^\mu dx^\nu,
\end{align}
where the perturbation part of the metric is expanded in a power series of  $\lambda$ and $\rho$:
\begin{align}
g_{\mu\nu}^\mathrm{(pert.)} 
= \sum_{n=0}^\infty g_{\mu\nu}^{(n)}(\rho,\tau,\theta,\varphi) \, \lambda^n
= \sum_{n=0}^\infty \sum_{m=0}^\infty g_{\mu\nu}^{(n,m)} (\tau,\theta,\varphi) \, \lambda^n \, \rho^m .
\end{align}
The vacuum Einstein equation is $\mathcal{R}_{\mu\nu} = 0$.
Here and henceforth, Greek indices are reserved for four-dimensional spacetime $(\rho, \tau, \theta, \varphi)$
and Latin ones for the distorted 2-sphere $(\theta, \varphi)$.
$\mathcal{R}_{\mu\nu}$ represents the Ricci tensor in four dimensions, while $R_{ij}$ that in two dimensions.
Each component of the Ricci tensor is also expanded in a power series of  $\lambda$ and $\rho$ in the same fashion as that of the metric
\begin{equation}
\mathcal{R}_{\mu\nu} = \sum_{n} \mathcal{R}_{\mu\nu}^{(n)} \, \lambda^n
 = \sum_n \sum_m \mathcal{R}_{\mu\nu}^{(n,m)} \, \lambda^n \, \rho^m.  \label{17}
\end{equation}
$\mathcal{R}_{\mu\nu}^{(n)}$ refers to the Ricci tensor at order $\lambda^n$ and $\mathcal{R}_{\mu\nu}^{(n,m)}$  at order $\lambda^n \rho^m$.
We found that the integer $n$ in the summation of (\ref{17}) should start from $-1$.
All $\mathcal{R}_{\mu\nu}^{(n,m)}$ should vanish if the metric is to be a solution to the Einstein equation.
We analysed equations both at the leading order ($n=-1$) and the subleading ($n=0$) one.
We will identify $\mathcal{R}_{\tau\tau}^{(-1,0)} = 0$ as the incompressibility condition, while $\mathcal{R}_{\tau i}^{(0,0)} = 0$ as the NS equation.

To relate $g^{(n)}_{\mu\nu}$ to fluid variables $v^i$ and $P$ such that if $v^i$ and $P$ satisfy fluid equations, then $g^{(n)}_{\mu\nu}$ satisfy Einstein equations, we impose the following requirements.
\begin{enumerate}
\item We choose a gauge $g_{\rho\rho} = 0$ not only for the background field but also for the full gravitational field including perturbations.
\item The leading order of the perturbations $g_{\mu\nu}^{(n)}$ is $n=0$.
This is expected because the leading behavior of $g_{\tau\tau}$ is $\mathcal{O}(\lambda^{-1})$ after the shift (\ref{phishift}). 
\item The leading and subleading perturbations $g_{\mu\nu}^{(0)}$ and $g_{\mu\nu}^{(1)}$ are cubic polynomials of $\rho$. Actually, it turns out they can even be quadratic polynomials. 
\item We require that the metric should be smooth on the boundary $\Sigma_c$, and that the fluid equation be expressed covariantly in terms of covariant derivatives with respect to the metric on the distorted 2-sphere (\ref{distortedsphere}). It turns out this requirement is strong enough to restrict the form of the fluid equation. 
We do not require that the induced metric on $\Sigma_c$ to coincide with or be conformal to that of a semi-direct sum of  time and the distorted sphere. 

\item We do not require that the perturbation part of the metric be written in terms of  covariant derivatives associated with the distorted sphere.
\end{enumerate}

Through analysis of the Einstein equations we found that by setting 
\begin{equation}
g^{(0,0)}_{\tau i}=c_0 \, v_i(\tau,\theta,\varphi), \label{incompressible}
\end{equation}
where $v_i$ is the velocity field and $c_0$ is a constant, it is possible to identify $\mathcal{R}_{\tau\tau}^{(-1,0)} = 0$ as the incompressibility condition.
\begin{equation}
\nabla_i \, v^i=0
\end{equation}Here $\nabla_i$ is covariant derivative associated with the metric on the distorted $S^2$. 
By solving other $\mathcal{R}_{\mu \nu}^{(n,m)} = 0$ it was also found that $\mathcal{R}_{\tau i}^{(0,0)} = 0$ yields the NS equation on the distorted $S^2$.
\begin{align}
\mathcal{R}_{\tau i}^{(0,0)} = & \frac{2c_2}{c_0} \partial_\tau v_i 
 - 2c_2 \, v^j \nabla_j v_i + c_1 \nabla_i P - \frac{c_0}{2} \left( \nabla^2 v_i + R_{ij} v^j \right) 
 -\frac{4c_2\pi^2 T_R^2}{c_0} \, \nabla_\varphi v_i\nonumber \\
& -\frac{8\cos{\theta}}{M^2c_0(3+\cos{2\theta})^3}
\left( 24c_2 M^2 \pi^2 T_R^2 + 5 c_0^2 + (8c_2 M^2 \pi^2 T_R^2 - c_0^2) \cos{2\theta} \right) \epsilon_{ij} v^j \nonumber \\
& = 0, \label{NS_eqns}
\end{align}
where $P(\tau,\theta,\varphi)$ is pressure, $c_1$ and $c_2$ are constants, $\epsilon_{ij}$ is a unit anti-symmetric tensor defined in (\ref{LeviCivita}). The term $R_{ij} \, v^j$ also appeared in the fluid equation dual to Schwartzschild black hole.\cite{Str1}
The appearance of  $\epsilon_{ij}$ in the last line may be surprising. However, a rotation about an axis breaks parity.\footnote{In \cite{G3}, \cite{G1}  a DC thermoelectric conductivity of field theory was considered within the context of AdS/CFT correspondence, and it was shown that this conductivity can be obtained by solving a system of generalized Stokes equations on perturbed black hole horizons.  We are informed that the term $v^j \, d\, \chi_{ji}^{(0)}$ in the Stokes equation (3.1) of \cite{G3} is similar to the last term proportional to $\epsilon_{ij} v^j$ in the middle of (\ref{NS_eqns}). } 
 The term proportional to $\nabla_{\varphi} \, v_i$ in (\ref{NS_eqns}) is also consistent with the isometry $\partial_{\varphi}$ of the background. 
Actually, to the order we are considering there also appears in NS equation another new term proportional to $\epsilon_{ij} \, \partial_{\tau} \, v^j$, which is consistent with the axial symmetry. This is, however, a result of the analysis to the leading and sub-leading orders. By analysis to higher orders,  we find that this term is excluded at the end. Hence the structure of the  Lagrange derivative of NS equation is not modified. 

The perturbative part of metric tensor which corresponds to this fluid equation is presented in appendix B. 
In the NS equation we can take $T_R \rightarrow 0$ limit. However, such a limit cannot be taken in the metric presented in appendix B. 
In (\ref{NS_eqns}) $\cos \theta$ and $\cos 2\theta$ can be re-expressed in terms of scalar curvature $R$ in (\ref{RR}). Hence this equation is covariant except for the presence of the term proportional to $\nabla_\varphi v_i$, which is also consistent with the symmetry. 

The explicit form of the fluid equation (\ref{NS_eqns}) is obtained by requirement 4 above. 
Equation ${\cal R}^{(0,0)}_{\tau i}=0$ $(i=\theta,\varphi)$ depends only on the metric components $g^{(0,0)}_{\tau i}$, $g^{(0,1)}_{\tau i}$, $g^{(1,0)}_{\rho i}$, $g^{(1,0)}_{\rho \tau} $ and  $g^{(0,1)}_{\tau\tau}$. 
Equation ${\cal R}^{(0,0)}_{\tau i}=0$  after substitution of (\ref{incompressible}) and ansatz\footnote{This is a generalization of the metric perturbation for Schwartzscild black hole.}
\begin{eqnarray}
g^{(0,1)}_{\tau i} &=& \alpha^{(1)}_{ij}(\theta) v^j, \nonumber \\
g^{(1,0)}_{\rho i} &=& \alpha^{(2)}_{ij}(\theta)v^j, \nonumber \\
g^{(1,0)}_{\rho \tau} &=& \frac{M^2 \Gamma}{\pi T_R} \, \Big(\beta^{(1)}(\theta)v^2+\beta^{(2)}(\theta)P\Big), \nonumber \\
g^{(0,1)}_{\tau\tau} &=& 4M^2 \Gamma \Big( \beta^{(3)}(\theta)v^2+\beta^{(4)}(\theta)P\Big), \label{ans}
\end{eqnarray}
where $\alpha^{(n)}(\theta)$ and $\beta^{(n)}(\theta)$ are unknown functions of $\theta$, takes a complicated structure. 
Especially, although NS equation must be a vector equation, terms such as $v_iv_j$, $v_i \nabla_jv_k$, $v^2$, $P$, $\partial_i P$, $v_i$, $\partial_{\tau}v_i$ and $\nabla_iv_j$ are included in this equation. To remove extra terms which cannot be combined into a vector quantity,
many conditions must be imposed on  $\alpha^{(n)}(\theta)$ and $\beta^{(n)}(\theta)$. Some of them are differential equations and solving them gives  the integration constants $c_n$ and determines $\alpha^{(n)}(\theta)$, $\beta^{(n)}(\theta)$ except for $\beta^{(2)}(\theta)$ and $\beta^{(3)}(\theta)$. Hence it turns out the requirement that the generalized NS equation be covariant (requirement 4.) provides sufficient constraints as strong as, if any, the boundary conditions on $\Sigma_c$.
\footnote{It is clear from (\ref{incompressible}), the first eq of (\ref{ans}) and (\ref{sol_ti1}) that we cannot impose a boundary condition that $g_{\tau i}^{(0)} = 0$ at $\rho = 1$.}
Eq (\ref{NS_eqns}) is the most general form, and no new terms appear.\footnote{If terms which are higher orders in $v_j$ were added to (\ref{ans}), then higher order terms of fluid equation would appear.} 
The undetermined functions $\beta^{(2)}(\theta)$, $\beta^{(3)}(\theta)$ do not appear in  (\ref{NS_eqns}) but appears in the metric. Here, it is crucial that further arbitrary terms such as $\nabla_i v^2$ and $\epsilon^{jk}(\nabla_j v_k)v_i$ cannot be introduced into the generalized NS equation by adding extra terms to the perturbation part of the metric tensor. 
By using the same prescription of this subsection, NS equation dual to Schwarzschild black hole can also be obtained. The result is presented in (\ref{SchwNS}) of Appendix C, which coincides with that of \cite{Str1}.

\subsection{Stationary solution to Navier-Stokes equation on the distorted 2-sphere}
\hspace{5mm}
If one fixes the constants of integration as
\begin{equation}
c_0 = -1, \quad c_1= -\frac{1}{2\nu}, \quad c_2 = \frac{1}{4\nu}, \label{fix_consts}
\end{equation}
 NS equation is obtained with the conventional weight
\begin{align}
  & \partial_\tau v_i + v^j \nabla_j v_i + \nabla_i P
  - \nu \left( \nabla^2 v_i + R_{ij} v^j \right)      
     -2 \pi^2 T_R^2 \nabla_\varphi v_i  \notag \\
 & + \frac{\cos{\theta}}{4 M^2 \Gamma^3}
   \left(6M^2 \pi^2 T_R^2 + 5 \nu + \left( 2 M^2 \pi^2 T_R^2 - \nu \right) \cos{2\theta} \right) \epsilon_{ij} v^j = 0,
\end{align}
where $\nu$ is a viscosity parameter. 

Kerr spacetime has isometry $\varphi \to \varphi + \mathrm{const.}$
Hence we expect a stationary solution $\bar{v}^i = (0,\xi)$, where $\xi$ is a constant, {\it i.e.},
\begin{equation}
\bar{v}_i = \xi \gamma_{ij} \bar{v}^j = 
	\begin{pmatrix}
		\xi \gamma_{\theta\theta} v^\theta \\ \xi \gamma_{\varphi\varphi} v^\varphi
	\end{pmatrix}
	 =
	 \begin{pmatrix}
	 	0 \\ \xi \gamma_{\varphi\varphi}
	 \end{pmatrix}
	 =
	 \begin{pmatrix}
	 	0 \\ \displaystyle{ \frac{2M^2 \xi \sin^2{\theta}}{\Gamma} } \label{xi}
	 \end{pmatrix}
\end{equation}
Substituting (\ref{xi}) into NS equation (\ref{NS_eqns}) yields two conditions.
\begin{align}
& \frac{\xi \left( (2-\Gamma) \nu + M^2 \Gamma \xi \right) \sin{2\theta}}{\Gamma^3}
- \partial_\theta P(\tau,\theta,\varphi) = 0,  \\
& \partial_\varphi P(\tau,\theta,\varphi) = 0. 
\end{align}
The solution for the pressure $P$ is given by
\begin{equation}
P(\tau,\theta,\varphi) = c_3 + \frac{2 \xi \left( (1-\Gamma) \nu + M^2 \Gamma \xi \right)}{\Gamma^2}
, \label{sol_stat}
\end{equation}
where $c_3$ is an integration constant. 
Choosing $c_3 = -2M^2 \xi^2$ reduces (\ref{sol_stat}) to 
\begin{equation}
P(\tau,\theta,\varphi) = \frac{1}{2} \bar{v}^2 + \frac{\nu}{2M^2 \Gamma} \bar{v}_\varphi.
\end{equation}

\subsection{Higher order equations}
\hspace*{5mm}
In appendix B some components of $g^{(n,m)}_{\mu\nu}$ are not presented. 
To determine $g_{\tau\tau}^{(0,0)}$ and $g_{ij}^{(1,0)}$ and $g_{\mu\nu}^{(2)}$ we need to solve 
 Einstein equations up to $\mathcal{O}(\lambda^0)$, {\em i.e.}, 
$\mathcal{R}_{\tau \tau}^{(0,m)}=0 \quad \mathrm{for~} m = 0, \cdots, 4$.

Equation $\mathcal{R}_{\tau\tau}^{(0,0)} = 0$ determines  a divergence of $g_{\tau i}^{(1,0)}$, defined by  
\begin{equation}
\nabla^i g_{\tau i}^{(1,0)} \equiv \frac{(\Gamma-2) \cot{\theta}}{2M^2 \Gamma^2} g_{\tau \theta}^{(1,0)}
+ \frac{1}{2M^2 \Gamma} \partial_\theta \, g_{\tau \theta}^{(1,0)}
+ \frac{\Gamma}{2M^2 \sin^2{\theta}} \partial_\varphi \, g_{\tau \varphi}^{(1,0)}.
\end{equation}
We have found that the equation takes a form
\begin{equation}
\mathcal{R}_{\tau\tau}^{(0,0)} = - 2 \pi T_R \nabla^i g_{\tau i}^{(1,0)} 
+ \mathcal{G} \left( g_{\tau\tau}^{(0,0)}, g_{ij}^{(1,0)}, v_i, P \right) = 0, \label{Rtt002a}
  \end{equation}
where $\mathcal{G}$ is some known function.
To solve the equation (\ref{Rtt002a}), we require that $\mathcal{G}$ be written in a form of  total divergence.
We accomplished this by tuning the form of $g_{\tau\tau}^{(0,0)}$ and $g_{ij}^{(1,0)}$ appropriately.
We set the following ansatz
\begin{align}
g_{\tau\tau}^{(0,0)} &= 0, \notag \\
g_{\theta\theta}^{(1,0)} &= \frac{c_1 M^2 \Gamma}{\pi T_R} P + \frac{c_2}{\pi T_R} v_\theta^2
-\frac{c_0^2 + 64 c_2 M^2 \pi^2 T_R^2 \Gamma (\Gamma-1)}{\pi T_R c_0 \sin^2{\theta}} v_\varphi
-\frac{2c_2 \Gamma^2}{\pi T_R \sin^2{\theta}} v_\varphi^2, \notag \\
g_{\theta\varphi}^{(1,0)} &= \frac{3c_2}{\pi T_R} v_\theta v_\varphi
 + \frac{3 c_0}{2\pi T_R} \partial_\varphi v_\theta, \notag \\
g_{\varphi\varphi}^{(1,0)} &= -\frac{c_1 M^2 \sin^2{\theta}}{\pi T_R \Gamma} P
-\frac{2c_2 \sin^2{\theta}}{\pi T_R \Gamma^2} v_\theta^2
+ \frac{c_0}{\pi T_R \Gamma^2} v_\varphi
+ \frac{c_2}{\pi T_R} v_\varphi^2,
\end{align}
and incompressibility condition (\ref{incompressible}) and the NS equation (\ref{NS_eqns}) makes $\mathcal{G}$ take
the form
\begin{align}
\mathcal{G} \left( g_{\tau\tau}^{(0,0)}, g_{ij}^{(1,0)}, v_i, P \right) &= \nabla^i V_i \notag \\
&= \frac{(\Gamma-2) \cot{\theta}}{2M^2 \Gamma^2} V_\theta
+ \frac{1}{2M^2 \Gamma} \partial_\theta \, V_\theta
+ \frac{\Gamma}{2M^2 \sin^2{\theta}} \partial_\varphi \, V_\varphi.
\end{align}
where
\begin{align}
V_\theta = & 2 c_1 c_0 P v_\theta + \frac{c_0^2 \sin{2\theta}}{8 M^2 \Gamma^2} v_\theta^2
+ \left( 36 c_2 \pi^2 T_R^2 - \frac{c_0^2}{M^2 \Gamma \sin^2{\theta}} \right) v_\theta v_\varphi \nonumber \\
& + \frac{c_0^2 \cot{\theta} (1+6 \csc^2{\theta} )}{4M^2} v_\varphi^2
+ \frac{c_0^2 \, \Gamma}{4M^2 \sin^2{\theta}} v_\varphi \nabla_\theta v_\varphi
+ \frac{5c_0^2 \, \Gamma}{4M^2 \sin^2{\theta}} v_\varphi \nabla_\varphi v_\theta \nonumber \\
& - \frac{\pi^2 T_R^2 c_0 ( 15 + \cos{4\theta})}{4 \Gamma^2} v_\theta
- \frac{2\pi^2 T_R^2 c_0 (2 \Gamma - 1) \cot{\theta}}{\Gamma} v_\varphi
+ 6 \pi^2 T_R^2 c_0 \nabla_\theta v_\varphi,
\end{align}
\begin{align}
V_\varphi = & \frac{ ( c_0^2 - 32 c_2 M^2 \pi^2 T_R^2 \Gamma) \sin^2{\theta}}{4M^2 \Gamma^3} v_\theta^2
+ \frac{3c_0^2 \cot{\theta}}{2M^2 \Gamma^2} v_\theta v_\varphi
-\frac{c_0^2}{4M^2\Gamma} v_\theta \nabla_\theta v_\varphi \nonumber \\
& + \frac{7 c_0^2}{4M^2 \Gamma} v_\theta \nabla_\varphi v_\theta
+ \left( -4 c_2 \pi^2 T_R^2 
+ \frac{3c_0^2}{4M^2 \Gamma} + \frac{c_0^2}{M^2 \sin^2{\theta}} \right) v_\varphi^2 \nonumber \\
& - \frac{20 c_1 M^2 \pi^2 T_R^2 \sin^2{\theta}}{\Gamma} P
+ \frac{4\pi^2 T_R^2 c_0 \sin{2\theta}}{\Gamma^3} v_\theta.
\end{align}
Now the equation (\ref{Rtt002a}) can be written as
\begin{align}
\mathcal{R}^{(0,0)}_{\tau\tau} = \nabla^i \left(V_i - 2\pi T_R \, g_{\tau i}^{(1,0)} \right) = 0 \label{Rtt002}
\end{align}
By choosing
\begin{equation}
g_{\tau i}^{(1,0)} = \frac{1}{2 \pi T_R} V_i, \label{326}
\end{equation}
 $\mathcal{R}_{\tau\tau}^{(0,0)} = 0$ is satisfied.

This result is quite different from that in \cite{Str1} for the study of dual fluid of Schwarzschild black hole.
In that case, time derivatives of $v^2$ and $P$ appear in the equation, $\mathcal{R}_{\tau\tau}^{(0,0)} = 0$, which determines $g_{\tau i}^{(1,0)}=\phi_i$,  and one has to remove the zero mode of $\partial_t P +\partial_t \, v^2/2$ on $S^2$ by shifting $P$ by some integral over $S^2$. 
In (\ref{Rtt002}), however, time derivatives of $v^2$ and $P$ do not appear. This is because we did not require the perturbation part of the metric be written in terms of covariant derivatives.  Similar prescription for Schwarzschild  black hole leads to a new metric which is dual to NS fluid. See appendix C. 

When the constants are fixed as in (\ref{fix_consts}),  $V_i$ are given by
\begin{align}
V_\theta = & -P v_\theta + \frac{\sin{2\theta}}{8 M^2 \Gamma^2} v_\theta^2
+ \left( -9 \pi^2 T_R^2 - \frac{1}{M^2 \Gamma \sin^2{\theta}} \right) v_\theta v_\varphi \nonumber \\
& + \frac{\cot{\theta} (1+6 \csc^2{\theta} )}{4M^2} v_\varphi^2
+ \frac{\Gamma}{4M^2 \sin^2{\theta}} v_\varphi \nabla_\theta v_\varphi
+ \frac{5 \Gamma}{4M^2 \sin^2{\theta}} v_\varphi \nabla_\varphi v_\theta \nonumber \\
& + \frac{\pi^2 T_R^2 ( 15 + \cos{4\theta})}{4 \Gamma^2} v_\theta
+ \frac{2\pi^2 T_R^2 (2 \Gamma - 1) \cot{\theta}}{\Gamma} v_\varphi
- 6 \pi^2 T_R^2 \nabla_\theta v_\varphi, \label{Vtheta}
\end{align}
\begin{align}
V_\varphi = & \frac{ ( 1 - 32 c_2 M^2 \pi^2 T_R^2 \Gamma) \sin^2{\theta}}{4M^2 \Gamma^3} v_\theta^2
+ \frac{3 \cot{\theta}}{2M^2 \Gamma^2} v_\theta v_\varphi
-\frac{1}{4M^2\Gamma} v_\theta \nabla_\theta v_\varphi \nonumber \\
& + \frac{7}{4M^2 \Gamma} v_\theta \nabla_\varphi v_\theta
+ \left( \pi^2 T_R^2 
+ \frac{3}{4M^2 \Gamma} + \frac{1}{M^2 \sin^2{\theta}} \right) v_\varphi^2 \nonumber \\
& - \frac{10 M^2 \pi^2 T_R^2 \sin^2{\theta}}{\Gamma} P
- \frac{4\pi^2 T_R^2 \sin{2\theta}}{\Gamma^3} v_\theta. \label{Vphi}
\end{align}
In each equations $\mathcal{R}_{\tau\tau}^{(0,m)} = 0, \quad (m=1,2,3)$
a non-derivative linear term $g_{\rho \tau}^{(2,m)}$ appears and we can easily solve the equations for them. 
The results are not presented in this paper. 
The equations $\mathcal{R}_{\theta\theta}^{(1,3)}$, $\mathcal{R}_{\varphi \varphi}^{(1,3)}$ and $\mathcal{R}_{\tau\tau}^{(0,4)}$ require $g_{\rho i}^{(2,3)} = 0$ and $g_{\rho \tau}^{(1,2)}=0$.
In this way the Einstein equation is solved up to  $\mathcal{O}(\lambda^0)$.

\section{Singularity of the Solution to the Fluid Equation}
\hspace*{5mm}
In this section we will consider a fluctuating event horizon ${\cal H}$. Its null normal is denoted as $k^{\mu}$. This normal vector defines a family of null geodesics on ${\cal H}$,  and by extending it off ${\cal H}$ appropriately, we will obtain an expansion scalar $\theta$ of null geodesics. This represents a logarithmic derivative of local area on ${\cal H}$, and if this is a true event horizon, the second law of black hole thermodynamics\cite{2ndlaw} asserts that  $\theta$ is not negative on the future horizon. 

In the case of the non-stationary black hole there are some proposals for definition of the  event horizon\cite{BHL}. In this paper we define the event horizon of a perturbed black hole around a stationary black hole as a null surface ${\cal H}$ which coincides with the event horizon of the unperturbed black hole when the gravitational perturbation is turned off. 
The focusing theorem implies that if $\theta$ is negative at some point on ${\cal H}$, then $\theta$ becomes $- \infty$ within a finite time and the null geodesics meet at caustic points. Hence if $\theta$ is not positive semi-definite, the horizon ${\cal H}$ defined above may not be a true event horizon in the light of the area theorem.  
The purpose of this section, however, is to study the singularity of the solution to the fluid equation, and we will continue to use ${\cal H}$ and study $\theta$ associated with ${\cal H}$. 
If $\theta$ is negative at some point, to avoid a  singularity of geodesics we need to impose an initial condition on the fluid velocity $v^i$ at some moment in the past and make $\theta$ take non-negative values at that time. Then  $\theta$ will be  positive or zero at later times and the solution to the NS equation has no singularity in the future.

 \subsection{Congruence of null geodesics}
\hspace*{5mm}
Let the horizon  ${\cal H}$ be a hypersurface $\Phi (x)=0$. 
The normal vector derived from $\Phi$,
\begin{eqnarray}
k_{\mu} = \partial_{\mu} \, \Phi
\end{eqnarray}
is null ($k^2=0$) on ${\cal H}$, but in general not null off ${\cal H}$. 
To compute the expansion scalar $\theta$ associated with a congruence of null geodesics, it is convenient to extend the null normal vector $k^{\mu}$ off the horizon ${\cal H}$.\footnote{A method different from that below is used in \cite{Carter}, \cite{G} .} This procedure will be briefly explained. We introduce another normal vector $N_{\mu}$ which satisfies the following conditions off ${\cal H}$ 
\begin{equation}
N^2=0, \qquad N \cdot k= -1
\end{equation}
and define
\begin{eqnarray}
\hat{k}_{\mu} &=& k_{\mu} +\frac{1}{2} \, \alpha \, N_{\mu} \, \Phi.
\end{eqnarray}
Here $\alpha$ is a function defined by 
\begin{equation}
k ^2= \alpha \, \Phi.
\end{equation}
Now the vector $\hat{k}_{\mu}$ is null in the neighborhood of ${\cal H}$ and defines a congruence of null geodesics on and off ${\cal H}$. The vector $\hat{k}_{\mu}$ is, however, hypersurface orthogonal only on ${\cal H}$. A projection matrix onto the subspace spanned by $\hat{k}_{\mu}$ and $N_{\mu}$ are defined by
\begin{equation}
h^{\mu\nu} = g^{\mu\nu} +\hat{k}^{\mu}N^{\nu}+N^{\mu} \hat{k}^{\nu}.
\end{equation}
Now the following tensor is introduced.\footnote{For review of congruence of geodesics, see \cite{Poisson}.}
\begin{eqnarray}
\tilde{B}_{\mu\nu}&=&{h_{\mu}}^{\lambda} \, {h_{\nu}}^{\rho}\, \nabla_{\rho} \, \hat{k}_{\lambda} \\
&=& \frac{1}{2} \, \theta \, h_{\mu\nu} +\sigma_{(\mu\nu)}+\omega_{[\mu\nu]} 
\end{eqnarray}
It can be shown that although the null vector $\hat{k}_{\mu}$ is not hypersurface orthogonal off ${\cal H}$, the rotation tensor $\omega_{\mu\nu}$ vanishes on ${\cal H}$. 

The expansion scalar 
\begin{eqnarray}
\theta &=& h^{\mu\nu} \, \tilde{B}_{\mu\nu}
\end{eqnarray}
obeys the Raychaudhuri's equation.
\begin{eqnarray}
\frac{d \theta}{d \mu} &=& \kappa \, \theta -\frac{1}{2} \, \theta^2 -\sigma^{\mu\nu} \, \sigma_{\mu\nu}+\omega^{\mu\nu} \, \omega_{\mu\nu}-R_{\mu\nu} \, \hat{k}^{\mu} \hat{k}^{\nu},
\label{Ray}
\end{eqnarray}
where $\mu$ parametrizes  null geodesics, $\hat{k}^{\nu}=d x^{\nu}(\mu)/d\mu$, and $\kappa (\mu)$ is a function defined by
\begin{eqnarray}
\hat{k}^{\nu} \, \nabla_{\nu} \, \hat{k}^{\mu} = \kappa(\mu) \, \hat{k}^{\mu}.
\end{eqnarray}

A special parametrization for which $\kappa=0$ is called an affine parametrization and this is achieved by a change of parametrization $\mu \rightarrow \chi$ (affine parameter) with $d\chi/d\mu=\exp (\int^{\mu} \kappa(\mu') d\mu')$.  Then $\theta$ is replaced by $\bar{\theta}=\theta \, \exp (-\int^{\mu} \kappa(\mu') d\mu')$ and $\bar{\theta}$ obeys  (\ref{Ray}) with $\kappa=0$. If matter stress energy tensor satisfies the null energy condition ($\hat{k}^{\mu} \, \hat{k}^{\nu} \, T_{\mu\nu} \geq 0$), then $R_{\mu\nu} \, \hat{k}^{\mu} \hat{k}^{\nu}  \geq 0$ owing to Einstein equation, and because $\omega_{\mu\nu}=0$ on ${\cal H}$, $\bar{\theta}$ satisfies the focusing theorem on ${\cal H}$.
\begin{eqnarray}
\frac{d \bar{\theta}}{d\chi} &\leq & -\frac{1}{2} \, \bar{\theta}^2
\end{eqnarray}
Hence if initially $\bar{\theta}=\bar{\theta}_0 <0$ at $\chi=\chi_0$, then $\bar{\theta} \rightarrow -\infty$ within finite affine parameter $\chi-\chi_0 \leq 2/|\bar{\theta}_0|$. This signals the occurrence of a caustic. 
For the singularity in $\bar{\theta}$ to be absent $\bar{\theta}$ must be non-negative. 
Because $\theta$ and $\bar{\theta}$ are related by a positive multiplicative factor, we will deal with $\theta$ in what follows. 

\subsection{Sign of expansion scalar $\theta$}
\hspace*{5mm}
When the perturbations in near-NHEK are absent, the event horizon is located at $\rho=0$. When the perturbations are switched on, the horizon is deformed. \cite{BHL}
\begin{eqnarray}
\Phi (\rho, \tau, \theta,\varphi) \equiv \rho-\lambda \, F_1(\tau, \theta,\varphi) -\lambda^2 \, F_2(\tau, \theta,\varphi)-\lambda^3 \, F_3(\tau, \theta,\varphi)+\cdots=0, \label{horizon}
\end{eqnarray}
where $F_a$ are unknown functions to be determined by a null normal condition, whose solution defines a new surface ${\cal H}$. 
 If $F_a$ are chosen such that $k \cdot k$ vanishes, then ${\cal H}$ becomes a null surface.\footnote{This procedure to define an event horizon is in the same spirit as in \cite{BHL}. This surface, however, may turn out not a true event horizon. 
In the discussion of this section, however,  it is not important whether our ${\cal H}$ is a true event horizon, or not.  } 
Generally, this procedure would yield a set of partial differential equations for $F_a$ and could not be solved explicitly. 

In the present case the condition of a null surface becomes algebraic equations for $F_1$ and $F_2$, because $g^{\rho \rho}$ takes a form $\frac{\rho}{\lambda}a(\tau, \theta,\varphi)+b(\tau, \theta,\varphi)+{\cal O}(\lambda)$, where $a(\tau, \theta,\varphi)$ and $b(\tau, \theta,\varphi)$ are known functions.  By substituting (\ref{horizon}) we have $g^{\rho \rho}=aF_1+b+\cdots$ and 
then equation $k^2=g^{\rho\rho}+{\cal O}(\lambda^1)=0$ is solved to yield 
\begin{eqnarray}
F_1 &=& -\frac{b}{a}=\frac{1}{16\pi M^4 \, T_R \Gamma^2} \, \Big[
-(g^{(0)}_{\tau\theta})^2-\Gamma^2 \, \csc^2\theta \, (g^{(0)}_{\tau\varphi})^2 \nonumber \\
&& \qquad \qquad +2M^2 \, \Gamma \, (g^{(0)}_{\tau\tau}-4\pi^2 \, T_R^2 \, g^{(0)}_{\tau\varphi})\Big]\Big|_{\rho=0}.
\end{eqnarray}
We checked that the induced metric on ${\cal H}$ is degenerate.

By using the procedure given in the previous subsection, $\theta$ is computed to ${\cal O}(\lambda^1)$. After using Einstein equation  we have $\theta=\theta^{(0)}+\lambda^1 \, \theta^{(1)}+{\cal O}(\lambda^2)$ with $\theta^{(0)}=0$ owing to incompressibility condition, and 
\begin{align}
\theta^{(1)} =& \frac{c_0^2}{8 M^6\pi T_R \Gamma^3} \, (\partial_{\theta}v_{\theta})^2
+\frac{c_0^2 \sin 2\theta}{16 M^6 \pi T_R \Gamma^4} \, v_{\theta} \, \partial_{\theta}v_{\theta} 
+\frac{c_0^2 \csc^2 \theta}{32 M^6 \pi T_R \Gamma} \, (\partial_{\varphi}v_{\theta})^2 \notag \\
& +\frac{c_0^2 \csc^2 \theta}{16 M^6\pi T_R \Gamma} \, \partial_{\varphi}v_{\theta} \partial_{\theta}v_{\varphi}
-\frac{c_0^2 \cot \theta \, \csc^2 \theta}{8 M^6 \pi T_R \Gamma^2} \, v_{\varphi} \partial_{\varphi}v_{\theta}
+\frac{c_0^2 \csc^2 \theta}{32 M^6 \pi T_R \Gamma} \, (\partial_{\theta}v_{\varphi})^2 \nonumber \\
& -\frac{c_0^2 \cot \theta \csc^2 \theta}{8 M^6 \pi T_R \Gamma^2} \, v_{\varphi} \partial_{\theta}v_{\varphi}
+\frac{c_0^2 \sin^2 2\theta}{128 M^6 \pi T_R \Gamma^5} \, (v_{\theta})^2 
+\frac{c_0^2\cot^2 \theta \csc^2 \theta}{8 M^6\pi T_R \Gamma^3} \, (v_{\varphi})^2 \label{theta1}
\end{align}
Although several terms are positive semi-definite, some are not. It must still be taken into account that in the above result the incompressibility condition (\ref{incompressible}) is not completely solved. In two spatial dimensions, however, the velocity of incompressible fluid can be expressed in terms of a scalar potential $f(\tau,\theta,\varphi)$:
\begin{equation}
v_i={\epsilon_{i}}^j \, \partial_j \, f.
\end{equation}
We checked that even when these equations are substituted into (\ref{theta1}), the result is still not positive semi-definite. 
The result (\ref{theta1}) does not depend on the ansatz (\ref{ans}) and the choice of $c_1$ and $c_2$.

The above result implies that in order to avoid the singularity of the expansion $\theta$ in the future, an initial condition on the velocity field $v_i$ must be imposed at some time $\tau=\tau_0$ such that $\theta^{(1)}(\tau_0,\theta,\varphi) \geq 0$ for all region of $\theta$ and $\varphi$ as well as condition of smoothness. Then the velocity field $v_i$ will remain non-singular afterwards.   If  $\theta^{(1)}(\tau_0,\theta,\varphi) \geq 0$ is not obeyed, however, caustics will occur on ${\cal H}$ in the future.  
We also performed similar analysis on the expansion scalar for the Schwarzschild black hole and obtained a same result. (See appendix C.)

The above result is relevant to  the existence and smoothness problem of the incompressible NS equation\cite{CMI}\cite{OzRabi}\cite{Str1}. Although this problem  is not solved in three spatial dimensions, it was studied in two dimensions (especially for $\boldmath{R}^2$ and $\boldmath{T}^2$)\cite{Lady}.  
Indeed, we can show that the expansion scalar is positive semi-definite in the case of a two-dimensional planar horizon in Rindler space (\ref{Rindler})\footnote{In the case of three-dimensional planar horizon, $\theta$ is also positive semi-definite. Surely a positive semi-definiteness of $\theta$ may not be a sufficient condition for existence of the smooth solution to incompressible NS equation.}. 
\begin{eqnarray}
\theta(t,x^1,x^2) &=& 4 \, (\partial_1\partial_2 \, f)^2+(\partial_1^2 \, f-\partial_2^2 \, f)^2 \geq 0, \nonumber \\
v_i(t, x^1,x^2) & \equiv & \epsilon_{ij} \, \partial_j \, f(t, x^1,x^2). \qquad (i,j=1,2)
\end{eqnarray}
However, solution of this problem in the case of a sphere is not known up to now.  Study of Fluid/Gravity correspondence along this line may shed some light on the solution of this problem
 for space with topologies other than plane. The above result is obtained  only for ${\cal O}(\lambda^1)$. It is interesting to investigate whether higher-order analysis imposes further restrictions on the initial conditions for $v^i$ to avoid caustics.

\section{Discussions}
\hspace*{5mm}
In this paper we derived an equation for viscous incompressible fluid which is dual to perturbations of near-NHEK black hole. This equation contains terms proportional to $\nabla_{\varphi} \, v_i$ and $\epsilon_{ij} \, v^j$, respectively. 
The structure of this hydrodynamic equation is interesting by itself and deserves further study. In this paper we considered only the near-NHEK geometry. This is simply because in the case of general Kerr geometry we did not succeed in finding appropriate expressions of the metric perturbation in terms of quantities in fluid dynamics such that if $v^i$ and $P$ satisfy the viscous NS equations, Einstein equation is also satisfied.  It is interesting to study whether this near horizon limit is essential, or there exist  dual fluids for any angular momentum parameter $a \, (\leq J)$. 
In this work we could not make metric perturbations to satisfy some boundary conditions on a cut-off surface $\Sigma_c$ except for smoothness.  Nonetheless the condition that the fluid equation can be expressed covariantly in terms of covariant derivatives with respect to the metric of the distorted 2-sphere is sufficient to restrict the form of the fluid equation.  It is also possible to derive a NS equation on the sphere dual to Schwarzschild black hole with only the requirement of subsec 3.2. (See appendix C.)

In sec. 4 singularity of the null geodesics on ${\cal H}$ (a null surface which agrees with the event horizon when the perturbation of the metric is turned off) in the future is studied. By computing the expansion scalar $\theta$ we found that for some choice of the initial condition on the velocity field, $\theta$ is negative in some region of the horizon and this causes the null geodesics to focus at caustic points in the future, and in turn, the singularity of the solution to the incompressible NS equations. To avoid such singularities it is necessary to set suitable initial conditions on the velocity at some time.

As for future studies an extension of the present work to the duality of fluid with higher dimensional rotating black hole and 
a study of turbulence based on the new fluid equation may be attempted. 
It is also interesting to study the possibility of systematic higher-order correction to the duality found in this paper.

\appendix

\section{Tensors on the distorted 2-sphere}
\label{2-dim tensor}\hspace{5mm}
The background metric on the distorted 2-sphere in the near-NHEK limit is
\begin{equation}
\gamma_{ij} =
	\begin{pmatrix}
		2 M^2 \Gamma(\theta) & 0 \\
		0 & \displaystyle{ \frac{2M^2 \sin^2{\theta}}{\Gamma(\theta)} }
	\end{pmatrix} \label{distortedsphere}
\end{equation}
where
\begin{equation}
\Gamma(\theta) \equiv \frac{1+\cos^2{\theta}}{2} = \frac{3 + \cos{2\theta}}{4}. 
\end{equation}
A covariant derivative associated with $\gamma_{ij}$ are denoted as $\nabla_i$. 
The epsilon tensor and the Ricci tensor are
\begin{equation}
\epsilon_{ij} \equiv \sqrt{| \det{\gamma_{ij}} |}
	\begin{pmatrix}
		0 & 1 \\
		-1 & 0
	\end{pmatrix}
= 2M^2 \sin{\theta}
	\begin{pmatrix}
		0 & 1 \\
		-1 & 0
	\end{pmatrix}, \label{LeviCivita}
\end{equation}
\begin{equation}
R_{ij} = -\frac{8(1+3\cos{2\theta})}{M^2(3+\cos{2\theta})^3} \, \gamma_{ij}
= -\frac{1+3\cos{2\theta}}{8M^2\Gamma^3} \, \gamma_{ij}.
\end{equation}
The Ricci scaler and the derivative are
\begin{equation}
 R = - \frac{1+3 \cos{2\theta}}{4 M^2 \Gamma^3}, \quad
 \partial_\theta R = \frac{3 \cos{\theta}\sin^3{\theta}}{M^2\Gamma^4}. \label{RR}
\end{equation}
The non-zero Christoffel symbols are
\begin{align}
\Gamma^\theta_{\theta\theta} = - \frac{\sin{2\theta}}{4\Gamma}, \quad
\Gamma^\theta_{\varphi\varphi} = - \frac{\sin{2\theta}}{2\Gamma^3}, \quad
\Gamma^\varphi_{\theta\varphi} = \frac{\cot{\theta}}{\Gamma}.
\end{align}

\section{Fluctuation of the metric tensor}
\hspace*{5mm}
We found the following results on the perturbations of the metric tensor,
\begin{align}
g_{\mu\nu}^\mathrm{(pert.)} 
= \sum_{n=0}^\infty g_{\mu\nu}^{(n)}(\rho,\tau,\theta,\varphi) \, \lambda^n
= \sum_{n=0}^\infty \sum_{m=0}^3 g_{\mu\nu}^{(n,m)} (\tau,\theta,\varphi) \, \lambda^n \, \rho^m, 
\end{align}
by an  expansion in a power series of  $\lambda$ and $\rho$:

\begin{equation}
g_{\rho\rho}^{(n,m)} = 0, \quad (\mathrm{for~} ^\forall n, ^\forall m) \label{sol_int}
\end{equation}
\begin{equation}
g_{\tau\tau}^{(0,3)} = 0,
\end{equation}
\begin{align}
g_{\tau\tau}^{(0,2)} = &
\frac{(-c_0^2+8c_2 M^2 \pi^2 T_R^2 \Gamma) \sin{2\theta}}{4\pi^2 T_R^2 c_0 \Gamma} v_\theta
+\frac{-c_0^2 + 8c_2 M^2 \pi^2 T_R^2 \Gamma+ c_0^2 \Gamma}{2 \pi^2 T_R^2 c_0} v_\varphi \notag \\
& +\frac{(-16 c_2 M^2 \pi^2 T_R^2 c_0^2 + c_0^4 + 128 c_2^2 M^4 \pi^4 T_R^4 \Gamma)\Gamma}{8 \pi^4 T_R^4 c_0^2} v^2
-2\pi T_R g_{\rho\tau}^{(1,1)} 
\end{align}
\begin{align}
g_{\tau\tau}^{(0,1)} = 4 M^2 \Gamma 
\left( \beta^{(3)} (\theta) v^2
+ \left( c_1- \beta^{(2)}(\theta) \right) P(\tau,\theta,\varphi) \right),
\end{align}
\begin{equation}
g_{\tau\tau}^{(0,0)} = 0,
\end{equation}
\begin{equation}
g_{\tau i}^{(0, m)} = 0, \quad (\mathrm{for~} m \geq 2 ) \label{sol_ti0}
\end{equation}
\begin{align}
g_{\tau i}^{(0,1)} & = \, \alpha^{(1)}_{ij} (\theta) \, v^j  \nonumber \\
& = -\frac{8 c_2 M^2 \Gamma}{c_0} \gamma_{ij}  v^j \nonumber \\
& = 
	\begin{pmatrix}
		\displaystyle{  -\frac{8 c_2 M^2 \Gamma}{c_0} }& 0 \\
		0 & \displaystyle{ - \frac{8c_2 M^2 \Gamma}{c_0} }
	\end{pmatrix}
	\begin{pmatrix}
		v_\theta \\ v_\varphi
	\end{pmatrix}, \label{sol_ti1}
\end{align}
\begin{align}
g_{\tau i}^{(0,0)} = c_0 \, v_i, \label{sol_ti2}
\end{align}
\begin{equation}
g_{\rho\tau}^{(1,3)} = 0,
\end{equation}
\begin{equation}
g_{\rho\tau}^{(1,2)} = 0,
\end{equation}
\begin{align}
g_{\rho\tau}^{(1,0)} = \frac{M^2 \Gamma}{\pi T_R} 
\left( \left( c_2 - \frac{2c_0^2}{M^2 \pi^2 T_R^2 \Gamma} -\beta^{(3)}(\theta) \right) v^2
+ \beta^{(2)}(\theta) P(\tau,\theta,\varphi) \right),
\end{align}
\begin{equation}
g_{\rho i}^{(1, m)} = 0, \quad (\mathrm{for~} m \geq 1 )
\end{equation}
\begin{align}
g_{\rho i}^{(1,0)} & = \, \alpha^{(2)}_{ij} (\theta) \, v^j \nonumber \\
& = \frac{-c_0^2 + 32 c_2 M^2 \pi^2 T_R^2 \Gamma}{8 \pi^3 T_R^3 c_0} \gamma_{ij}  v^j \nonumber \\
& = 
	\begin{pmatrix}
		\displaystyle{ \frac{-c_0^2 + 32 c_2 M^2 \pi^2 T_R^2 \Gamma}{8 \pi^3 T_R^3 c_0} }& 
		0 \\
		0 &
		\displaystyle{ \frac{-c_0^2 + 32 c_2 M^2 \pi^2 T_R^2 \Gamma}{8 \pi^3 T_R^3 c_0} }
	\end{pmatrix}
	\begin{pmatrix}
		v_\theta \\ v_\varphi
	\end{pmatrix}, \label{sol_ti2}
\end{align}
\begin{equation}
g_{\tau i}^{(1,3)} = 0,
\end{equation}
\begin{align}
g_{\tau \theta}^{(1,2)} = 
& \frac{g_{\rho \tau}^{(1,1)} \sin{2\theta}}{\Gamma} + \frac{1}{2} \partial_\theta \, g_{\rho \tau}^{(1,1)} \notag \\
& + \frac{c_2 \left( -16 c_2 M^2 \pi^2 T_R^2 c_0^2 + c_0^4 + 128 c_2^2 M^4 \pi^4 T_R^4 \Gamma \right)}
{16 M ^2 \pi^5 T_R^5 c_0^3 } v_\theta
\left( v_\theta^2 + \frac{\Gamma^2}{\sin^2{\theta}} v_\varphi^2 \right)
\notag \\
& + \frac{c_2 \left( - c_0^2 + 4 c_2 M^2 \pi^2 T_R^2 \Gamma \right) \sin{2\theta}}{4 \pi^3 T_R^3 c_0^2 \Gamma} v_\theta^2
+ \frac{c_2 \left(c_0^2(-1+\Gamma) + 8 c_2 M^2 \pi^2 T_R^2 \Gamma \right)}{2\pi^3 T_R^3 c_0^2} v_\theta v_\varphi \notag \\
& + \frac{-16 c_2 M^2 \pi^2 T_R^2 c_0^2 + c_0^4 + 128 c_2^2 M^4 \pi^4 T_R^4 \Gamma}{32 M^2 \pi^5 T_R^5 c_0^2}
 v_\theta \nabla_\theta v_\theta \notag \\
 & + \frac{c_2 \Gamma \cot{\theta}}{2 \pi^3 T_R^3} v_\theta \left( -\nabla_\theta v_\varphi + \nabla_\varphi v_\theta \right)
 -\frac{2 c_2^2 M^2 \Gamma^2 \cot{\theta}}{\pi T_R c_0^2} v_\varphi^2 \notag \\
 & + \frac{\Gamma^2 \left(-32 c_2 M^2 \pi^2 T_R^2 c_0^2 + c_0^4 + 384 c_2^2 M^4 \pi^4 T_R^4 \Gamma \right)}{32M^2 \pi^5 T_R^5 c_0^2 \sin^2{\theta}} v_\varphi \nabla_\theta v_\varphi \notag \\
 & - \frac{c_2 \Gamma^2 \left(- c_0^2 + 16 c_2 M^2 \pi^2 T_R^2 \Gamma \right)}{2 \pi^3 T_R^3 c_0^2 \sin^2{\theta}} v_\varphi \nabla_\varphi v_\theta \notag \\
 & + \frac{-8 c_2 M^2 \pi^2 T_R^2 (-1 + \Gamma) \Gamma + c_0^2 (-1 +4\Gamma -2 \Gamma^2)
 - 16 c_2 \pi^3 T_R^3 \Gamma^2 g_{\rho \tau }^{(1,1)}}{8 \pi^3 T_R^3 c_0 \Gamma^2} v_\theta \notag \\
 & + \frac{ \left(c_0^2 (1+\Gamma) + 8 c_2 M^2 \pi^2 T_R^2 \Gamma (-7 + 4 \Gamma) \right) \cot{\theta}}{8\pi^3 T_R^3 c_0 \Gamma} v_\varphi \notag \\
 & - \frac{\left( c_0^2 - 8c_2 M^2 \pi^2 T_R^2 \Gamma \right) \sin{2\theta}}{16\pi^3 T_R^3 c_0 \Gamma} \nabla_\theta v_\theta
+ \frac{c_0^2 (-1+ \Gamma) + 24 c_2 M^2 \pi^2 T_R^2 \Gamma}{8 \pi^3 T_R^3 c_0} \nabla_\theta v_\varphi \notag \\
& - \frac{2 c_2 M^2 \Gamma}{\pi T_R c_0} \nabla_\varphi v_\theta
+ \frac{c_2 M^2 \Gamma^3}{\pi T_R c_0 \sin^2{\theta}} \left( -\nabla_\theta \nabla_\varphi v_\varphi
+ \nabla_\varphi \nabla_\varphi v_\theta \right),
\end{align}
\begin{align}
g_{\tau \varphi}^{(1,2)} =
& + \frac{g_{\rho \tau}^{(1,1)} \sin^2{\theta}}{2\Gamma^2} + \frac{1}{2} \partial_\varphi g_{\rho\tau}^{(1,1)} \notag \\
& + \frac{c_2 \left( -16 c_2 M^2 \pi^2 T_R^2 c_0^2 + c_0^4+128 c_2^2 M^4 \pi^4 T_R^4 \Gamma \right)}{16M^2 \pi^5 T_R^5 c_0^3} v_\varphi
\left( v_\theta^2 + \frac{\Gamma^2}{\sin^2{\theta}} v_\varphi^2 \right) \notag \\
& - \frac{\left(-16 c_2 M^2 \pi^2 T_R^2 c_0^2 + c_0^4 + 128 c_2^2 M^4 \pi^4 T_R^4 \Gamma \right) \sin^2{\theta}}{64M^2 \pi^5 T_R^5 c_0^2 \Gamma^2} v_\theta^2 \notag \\
& + \frac{c_2 \left( -c_0^2 + 8c_2 M^2 \pi^2 T_R^2 \Gamma \right) \sin{2\theta}}{4 \pi^3 T_R^3 c_0^2 \Gamma} v_\theta v_\varphi 
+ \frac{c_2 \left( c_0^2 -16 c_2 M^2 \pi^2 T_R^2 \Gamma \right)}{2 \pi^3 T_R^3 c_0^2} v_\theta \nabla_\theta v_\varphi \notag \\
& + \frac{-32 c_2 M^2 \pi^2 T_R^2 c_0^2 + c_0^4 + 384 c_2^2 M^4 \pi^4 T_R^4 \Gamma}{32M^2 \pi^5 T_R^5 c_0^2} v_\theta \nabla_\varphi v_\theta \notag \\
 & + \frac{-c_0^4 + 128 c_2^2 M^4 \pi^4 T_R^4 \Gamma + 16 c_2 M^2 \pi^2 T_R^2 c_0^2 (-1 + 2\Gamma)}{64M^2 \pi^5 T_R^5 c_0^2} v_\varphi^2 \notag \\
& + \frac{c_2 \Gamma \cot{\theta}}{2 \pi^3 T_R^3} v_\varphi \left( - \nabla_\theta v_\varphi
+ \nabla_\varphi v_\theta \right) \notag \\
 & + \frac{\Gamma^2 \left(-16 c_2 M^2 \pi^2 T_R^2 c_0^2 + c_0^4 + 128 c_2^2 M^4 \pi^4 T_R^4 \Gamma \right)}{32 M^2 \pi^5 T_R^5 c_0^2 \sin^2{\theta}} v_\varphi \nabla_\varphi v_\varphi \notag \\
 & - \frac{ \left( c_0^2 (1+\Gamma) + 8 c_2 M^2 \pi^2 T_R^2 \Gamma (-5 + 2\Gamma) \right) \sin{2\theta}}{16 \pi^3 T_R^3 c_0 \Gamma^3} v_\theta \notag \\
 & + \frac{8c_2 M^2 \pi^2 T_R^2 \Gamma (1-2 \Gamma^2) + c_0^2 (-1+4\Gamma -2 \Gamma^2) - 16 c_2 \pi^3 T_R^3 \Gamma^2 g_{\rho\tau}^{(1,1)}}{8\pi^3 T_R^3 c_0 \Gamma^2} v_\varphi \notag \\
 & - \frac{ c_2 M^2 \sin{2\theta}}{\pi T_R c_0} \nabla_\theta v_\varphi
 - \frac{ \left( c_0^2 -24 c_2 M^2 \pi^2 T_R^2 \Gamma \right) \sin{2\theta}}{16 \pi^3 T_R^3 c_0 \Gamma} \nabla_\varphi v_\theta \notag \\
 & + \frac{c_0^2 (-1+ \Gamma) + 8 c_2 M^2 \pi^2 T_R^2 \Gamma}{8\pi^3 T_R^3 c_0} \nabla_\varphi v_\varphi
 + \frac{c_2 M^2 \Gamma}{\pi T_R c_0} \left( \nabla_\theta \nabla_\theta v_\varphi - \nabla_\theta \nabla_\varphi v_\theta \right),
\end{align}
\begin{equation}
g_{\tau i}^{(1,0)} = \frac{1}{2 \pi T_R} V_i,
\end{equation}
\begin{equation}
g_{ij}^{(1,m)} = 0, \quad (\mathrm{for~} m \geq 2 )
\end{equation}
\begin{align}
g_{ij}^{(1,1)} = &
-\frac{4c_2^2 M^2 \Gamma}{\pi T_R c_0^2} v_i v_j
+\left( -\frac{c_0}{8\pi^3 T_R^3} + \frac{2c_2 M^2 \Gamma}{\pi T_R c_0} \right) \nabla_i v_j
-\frac{c_0\cos{\theta}}{8\pi^3T_R^3} \nabla_i v^k \epsilon_{kj} \nonumber \\
& -\frac{2c_2 M^4 \Gamma^4}{3\pi T_R c_0 \sin^2{\theta}} v_i 
\left( \partial_j R + \frac{1}{\cos{\theta}} \epsilon_j{}^k \partial_k R \right)
- \frac{M^2 c_0 \Gamma^4}{24\pi^3 T_R^3 \sin^2{\theta} \cos{\theta}} \epsilon_{ik} v^k \partial_j R \nonumber \\
& + \left( i \leftrightarrow j \right),
\end{align}
\begin{align}
g_{\theta\theta}^{(1,0)} &= \frac{c_1 M^2 \Gamma}{\pi T_R} P + \frac{c_2}{\pi T_R} v_\theta^2
-\frac{c_0^2 + 64 c_2 M^2 \pi^2 T_R^2 \Gamma (\Gamma-1)}{\pi T_R c_0 \sin^2{\theta}} v_\varphi
-\frac{2c_2 \Gamma^2}{\pi T_R \sin^2{\theta}} v_\varphi^2, \notag \\
g_{\theta\varphi}^{(1,0)} &= \frac{3c_2}{\pi T_R} v_\theta v_\varphi
 + \frac{3 c_0}{2\pi T_R} \partial_\varphi v_\theta, \notag \\
g_{\varphi\varphi}^{(1,0)} &= -\frac{c_1 M^2 \sin^2{\theta}}{\pi T_R \Gamma} P
-\frac{2c_2 \sin^2{\theta}}{\pi T_R \Gamma^2} v_\theta^2
+ \frac{c_0}{\pi T_R \Gamma^2} v_\varphi
+ \frac{c_2}{\pi T_R} v_\varphi^2.
\end{align}
$\alpha^{(i)}(\theta)$ and $\beta^{(i)}(\theta)$ are undetermined functions of $\theta$ and $V_i$ is defined by (\ref{Vtheta}) and (\ref{Vphi}). 
The terms which do not appear above can be arbitrary functions up to the order we consider.

\section{Metric for the perturbation of Schwarzschild black hole and expansion scalar}
\hspace*{5mm}
We also derive NS equation for the perturbation to the Schwarzschild background.
The boundary condition on $\Sigma_c$ is only that the functions $v_i$ and $P$ are smooth.
Here we do not require that the perturbation part of the metric  be written covariantly in terms of covariant derivatives associated with the 2-sphere.

The background metric in a power series of $\lambda$ is
\begin{align}
ds^2_\mathrm{Sch} = 
2 dt d\rho
+ \left( - \frac{\rho}{2M \lambda} + \frac{\rho^2}{4M^2} - \frac{\rho^3 \lambda}{8M^3} \right) dt^2
+ \left( 1+ \rho \, \lambda \right) 4 M^2 d \Omega^2_2
+ \mathcal{O} (\lambda^2)
\end{align}
where M is a mass of the Schwarzschild black hole.

The perturbation of the metric is expressed as
\begin{align}
g_{\mu\nu}^\mathrm{(pert.)} 
= \sum_{n=0}^\infty g_{\mu\nu}^{(n)}(\rho,\tau,\theta,\varphi) \, \lambda^n
= \sum_{n=0}^\infty \sum_{m=0}^3 g_{\mu\nu}^{(n,m)} (\tau,\theta,\varphi) \, \lambda^n \, \rho^m. 
\end{align}
The non-vanishing coefficient functions are given as follows. 
\begin{align}
g_{\tau\tau}^{(0)} = - \frac{\rho}{\nu} P,
\end{align}
\begin{align}
g_{\tau i}^{(0)} = \left( -1 + \frac{\rho}{\nu} \right) v_i,
\end{align}
\begin{align}
g_{\rho \tau}^{(1)} = \frac{M}{\nu^2} \left( -3 \nu + \rho \right) v^2,
\end{align}
\begin{align}
g_{\tau \theta}^{(1,0)} = 
& \frac{2M}{\nu} P \, v_\theta 
+ \frac{\cot{\theta}}{2M} v_\theta^2
+ \frac{\csc^2{\theta}}{2M} v_\varphi \nabla_\theta v_\varphi
- \frac{\cot{\theta} \csc^2{\theta}}{2M} v_\varphi^2 \notag \\
& - \frac{\csc^2{\theta}}{M} v_\varphi \nabla_\varphi v_\theta
- \frac{\csc^2{\theta}}{2M} v_\theta \nabla_\varphi v_\varphi,
\end{align}
\begin{align}
g_{\tau \theta}^{(1,2)} = 
-\frac{1}{4M \nu} v_\theta
+ \frac{1}{2M \nu^2} v_\theta \nabla_\theta v_\theta
+ \frac{\csc^2{\theta}}{4M\nu} \nabla_\theta \nabla_\varphi v_\varphi
+ \frac{\csc^2{\theta}}{2M \nu^2} v_\varphi \nabla_\varphi v_\theta
- \frac{\csc^2{\theta}}{4M \nu} \nabla_\varphi^2 v_\theta,
\end{align}
\begin{align}
g_{\tau \varphi}^{(1,0)} =
\frac{2M}{\nu} P \, v_\varphi
+ \frac{1}{M} v_\varphi \nabla_\theta v_\theta
+ \frac{1}{2M} v_\theta \nabla_\varphi v_\theta
+ \frac{\csc^2{\theta}}{2M} v_\varphi \nabla_\varphi v_\varphi,
\end{align}
\begin{align}
g_{\tau \varphi}^{(1,2)} = 
- \frac{1}{4M \nu} \nabla_\theta^2 v_\varphi
+ \frac{1}{2M \nu^2} v_\theta \nabla_\theta v_\varphi
+ \frac{1}{4M \nu} \nabla_\theta \nabla_\varphi v_\theta
- \frac{1}{2M \nu} v_\varphi
+ \frac{\csc^2{\theta}}{2M \nu^2} v_\varphi \nabla_\varphi v_\varphi,
\end{align}
\begin{align}
g_{\theta\theta}^{(1)} = \frac{2M}{\nu} v_\theta^2 - \frac{4M \csc^2{\theta}}{\nu} v_\varphi^2
+\rho \left( - \frac{2M}{\nu^2} v_\theta^2 + \frac{4M}{\nu} \nabla_\theta v_\theta \right),
\end{align}
\begin{align}
g_{\theta \varphi}^{(1)} =
\frac{6M}{\nu} v_\theta v_\varphi
+ \frac{2M \rho}{\nu} \left( \nabla_\theta v_\varphi + \nabla_\varphi v_\theta - \frac{1}{\nu} v_\theta v_\varphi
 \right),
\end{align}
\begin{align}
g_{\varphi \varphi}^{(1)} = 
& - \frac{4M \sin^2{\theta}}{\nu} v_\theta^2
+ \frac{2M}{\nu} v_\varphi^2
+ 8 M \sin^2{\theta} \left( \nabla_\theta v_\theta
+ \csc^2{\theta} \nabla_\varphi v_\varphi \right) \notag \\
&
+ \rho \left( -\frac{2M}{\nu^2} v_\varphi^2 + \frac{4M}{\nu} \nabla_\varphi v_\varphi \right).
\end{align}
Like (\ref{326}) $g^{(1,0)}_{\tau i}$ is explicitly solved. 
This metric satisfies Einstein equation provided $v^i$ and $P$ satisfy an incompressibility condition
\begin{equation}
\mathcal{R}_{\tau\tau}^{(-1,0)} = \nabla^i v_i = 0,
\end{equation}
and NS equation
\begin{equation}
\mathcal{R}_{\tau i}^{(0,0)} = 
\partial_\tau v_i + v^j \nabla_j v_i + \nabla_i P - \nu \left( \nabla^2 v_i + R_{ij} v^j \right) = 0, \label{SchwNS}
\end{equation}
where $R_{ij}$ and $\nabla_i$ are Ricci tensor and covariant derivative associated with the distorted 2-sphere. It is checked that 
all other components of the Ricci tensor $\mathcal{R}_{\mu\nu}^{(-1)}$ and $\mathcal{R}_{\mu\nu}^{(0)}$ vanish 
except for $\mathcal{R}_{\mu\nu}^{(0,m)} ~ (m=1,2,3,4)$. In order to make these vanish $g^{(2,m)}$ must be adjusted. 

The expansion scalar $\theta$ with perturbation in the Schwarzschild black hole is given by
\begin{align}
\theta =
& \lambda \, \Big(\frac{\cot^2{\theta} \csc^2{\theta}}{2M^3} v_\varphi^2
- \frac{\cot{\theta} \csc^2{\theta}}{2M^3} v_\varphi \partial_\varphi v_\theta
- \frac{\cot{\theta} \csc^2{\theta}}{2M^3} v_\varphi \partial_\theta v_\varphi
+ \frac{\csc^2{\theta}}{8M^3} \left( \partial_\varphi v_\theta \right)^2 \notag \\
& + \frac{\csc^2{\theta}}{4M^3} \partial_\varphi v_\theta \, \partial_\theta v_\varphi
+ \frac{1}{2M^3} \left( \partial_\theta v_\theta \right)^2
+ \frac{\csc^2{\theta}}{8M^3} \left( \partial_\theta v_\varphi \right)^2 \Big) + \mathcal{O} (\lambda^2).
\end{align}
This is not positive semi-definite as in the case of Kerr black hole.
We checked that this conclusion does not change, even if the metric in \cite{Str1} is used.

\end{document}